\documentclass[pra,aps,reprint,superscriptaddress,twocolumn,longbibliography]{revtex4-1} 
\usepackage{graphicx,color,appendix,physics,wasysym}
\usepackage{setspace,appendix} 
\usepackage{amsmath,bm}
\usepackage{amssymb,braket}
\usepackage{hyperref}
\hypersetup{colorlinks,linkcolor={red},citecolor={green},urlcolor={blue}}

\newcommand{\bla}{\color{black}}

\usepackage{titlesec}
\usepackage[T1]{fontenc}

\begin{document}
	\title{Singularities, mixing and non-Markovianity of Pauli dynamical maps}
	\author{Shrikant Utagi}
	\email{shrikant.phys@gmail.com}
	\affiliation{Theoretical Sciences Division, Poornaprajna Institute of Scientific Research, Bidalur
		Bengaluru-- 562164, India}
	\affiliation{Graduate Studies, Manipal Academy of Higher Education, Manipal -576104, India.}
	\author{Vinod N. Rao}
	\affiliation{Theoretical Sciences Division, Poornaprajna Institute of Scientific Research, Bidalur
		Bengaluru-- 562164, India}
	\author{R. Srikanth}
	\email{srik@poornaprajna.org}
	\affiliation{Theoretical Sciences Division, Poornaprajna Institute of Scientific Research, Bidalur
		Bengaluru-- 562164, India}
	\author{Subhashish Banerjee.}
	\email{subhashish@iitj.ac.in}
	\affiliation{Interdisciplinary Program  on Quantum Information and Computation (IDRP-QIC),
		Indian Institute of Technology, Jodhpur-342037, India.}	
	
	\begin{abstract}
	Quantum non-Markovianity of channels can be produced by mixing Markovian channels, as observed  recently by various authors. We consider an analogous question of whether singularities of the channel can be produced by mixing non-singular channels, i.e., ones that lack them.  Here we answer the question in the negative in the context of qubit Pauli channels. On the other hand, mixing channels with a singularity can lead to the elimination of singularities in the resultant channel. We distinguish between two types of singular channels, which lead under mixing to broadly quite different properties of the singularity in the resultant channel. The connection to non-Markovianity (in the sense of completely positive indivisibility) is pointed out. These results impose nontrivial restrictions on the experimental realization of non-invertible quantum channels by a process of channel mixing.
	\end{abstract}
\maketitle

\section{Introduction}
Open quantum systems, which are systems in interaction with an ambient environment \cite{banerjee2018open}, experience an evolution with a rich structure showing the absence or presence of memory effects  \cite{breuer2002theory, srikanth2008squeezed, omkar2013dissipative, RHP10,breuer2009measure, hall2014canonical, shrikant2020temporal, li2019non}, unital or non-unital features \cite{verstraete2002quantum, ruskai2002analysis, banerjee2007dynamics, naikoo2020coherence}. Open system effects have profound ramifications in areas such as those in quantum thermodynamics \cite{esposito2009nonequilibrium, campisi2011colloquium, thomas2018thermodynamics}, quantum cryptography \cite{thapliyal2017quantum, shrikant2020ping}, quantum walks \cite{kumar2018non, naikoo2020non}, quantum correlations and coherence \cite{bhattacharya2018evolution, banerjee2010dynamics, naikoo2019facets}, among others (cf. \cite{li2020non}).

An open quantum system evolution, under quite general conditions, is known to be described by the general master equation 
\begin{align}
	\dot{\rho}(t) &= -\frac{i}{\hbar}[H_S(t),\rho(t)] + \sum_j \gamma_j(t)\left(L_j(t)\rho(t) L^{\dagger}_j(t) \right. \nonumber \\
	\quad \quad	& \left. - \frac{1}{2}\{L^{\dagger}_j(t) L_j(t),\rho(t)\}\right) \nonumber \\ & \equiv \mathcal{L}(t)[\rho(t)],
	\label{eq:lindblad-like}
\end{align}
where $\gamma_j(t)$ are the time-dependent decay rates, and $\{L_j\}$ are the set of orthonormal trace-less operators. The time-independent version of Eq. (\ref{eq:lindblad-like}), was studied in the pioneering works \cite{gorini1976completely, lindblad1975completely}.

Equivalently,  an open quantum system evolution can be described by a channel, i.e., a completely positive (CP) dynamical map, which is given by operator-sum (or Kraus) representation: 
\begin{align}
	\rho(t)=\mathcal{E}(t)[\rho(0)] = \sum_j K_j(t) \rho K^\dagger _j (t),
	\label{eq:kraus}
\end{align}
where $K_j(t)$ are the Kraus operators. The dynamical map $\mathcal{E}(t)$ itself obeys the master equation  \cite{chruscinski2010non,hall2014canonical} 
$
\mathcal{\dot{E}}(t) = \mathcal{L}(t)[ \mathcal{E}(t)],
$ so that
the time-dependent map has the solution \begin{align}
\mathcal{E}(t, t_i) = \mathcal{T} {\rm exp} \biggl\{ \int_{t_i}^{t} \mathcal{L}(s) ds  \biggr\}, 
\label{eq:mapintegral}
\end{align} for all  $t_i \le s \le t$, where $\mathcal{T}$ is the time-ordering operator. 
And furthermore, 
\begin{equation}
\mathcal{L}(t) = \mathcal{\dot{E}}(t) \mathcal{E}^{-1}(t),
\label{eq:E2L}
\end{equation}
showing that non-invertibility of the map $\mathcal{E}(t)$ corresponds to a singularity in the generator $\mathcal{L}(t)$.

 Complete positivity of $\mathcal{E}(t,t_i)$ is the requirement that not only is $\mathcal{E}(t,t_i)$ positive, but so is any extension $\mathcal{E}(t,t_i) \otimes \mathbb{I}_d$, where $\mathbb{I}_d$ is the identity operator in the Hilbert space of a $d$-dimensional ancilla. The map $\mathcal{E}(t,t_i)$ is CP if and only if the Choi matrix $\chi = (\mathcal{E}(t,t_i)\otimes I)[\vert \psi^+ \rangle \langle \psi^+ \vert] \ge 0$ for all $t \ge t_i$, where $\ket{\psi^+} = \ket{00}+\ket{11}$ is an unnormalized maximally entangled state. Consider the two-parameter composition of a CP map $\mathcal{E}(t_f,t_i)$ given by
 \begin{align}
 \mathcal{E}(t_f,t_i)=\mathcal{E}(t_f,t)\mathcal{E}(t,t_i).
 	\label{eq:2-parameter-law}
 \end{align}
If for all $t_f \ge t \ge t_i$, the intermediate map $\mathcal{E}(t_f,t)$ is CP, then the map $\mathcal{E}(t_f,t_i)$ is called CP-divisible \cite{RHP10,chruscinski2011divisiblity}. Otherwise, it is CP-indivisible. CP-indivisibility has been proposed as one of the criteria for non-Markovian evolution, among a plethora of others \cite{RHP14,breuer2009measure, chruscinski2014degree,bylicka2014non,li2018concepts,shrikant2020temporal, shrikant2021causality}. In this work, we consider the concept of non-Markovianity of a channel in the sense of CP-indivisibility, as defined above.

Important for our purpose, in this work, are the notions of singular points of the map and of singular channels.

\textit{Definition 1.} If there is a time $t=t_\ast$ such that the composition Eq. (\ref{eq:2-parameter-law}) fails, because the map $\mathcal{E}(t_\ast,t_i)$ is non-invertible and thus $\mathcal{E}(t_f,t_\ast) \equiv \mathcal{E}(t_f,t_i)\mathcal{E}(t_\ast,t_i)^{-1}$ is undefined, then the point $t_\ast$ is called the singularity (or, singular point) of the channel $\mathcal{E}(t_f,t_i)$ \cite{hou2012singularity}. Furthermore, the channel is called ``singular''. If no such singular points $t_\ast$ exist, then the channel is said to be non-singular (or, regular).

Note that the singularity of the channel can be accompanied by perfectly regular dynamics, i.e., the map $\mathcal{E}(t_\ast,t_i)$ itself is well-defined (albeit non-invertible) \cite{chruscinski2010non,shrikant2018non-Markovian}. 

Ref. \cite{hou2011alternative} discusses a method for making singularities tractable in the context of the definition of CP-divisibility of maps \cite{RHP10}. Building thereon, a measure of singularities of the maps is presented in \cite{hou2012singularity}. An account of handling the singularities was reviewed in \cite[Sec. 4.3]{RHP14}. These measures based on CP-indivisibility are equivalent to the one based on decay rate \cite{hall2014canonical} up to a constant factor.

 \bla

The effect of mixing different quantum evolutions has attracted attention of late.  References. \cite{wolf2008assessing,chruscinski2010long} show that a convex combination of semigroup dynamical maps can lead to a deviation from the semigroup structure. Quite interestingly, the convex combination $\mathcal{E}^\prime= (1-p)\mathcal{E}_1 + p \mathcal{E}_2 $ of two semigroup (hence, CP-divisible) maps $\mathcal{E}_1 = {\rm exp}\{t\mathcal{L}_1\}$ and $\mathcal{E}_2 = {\rm exp}(t\mathcal{L}_2)$ may give rise to CP-indivisible (even eternally CP-indivisible) evolution \cite{wudarski2016markovian}. 
More recently various authors have shown that it is possible to obtain a CP-indivisible Pauli channel by mixing CP-divisible Pauli channels \cite{wudarski2016markovian, jagadish2020convex, jagadish2020+convex}, implying that the set of CP-divisible channels is not convex.

For almost all relevant works in the literature, including those cited above \cite{hou2011alternative,hou2012singularity,RHP14,shrikant2018non-Markovian, shrikant2020quasi}, instances of singularity of a channel are always accompanied by non-Markovianity in the sense of CP-indivisibility (though the converse is not true). In this light, our above observation concerning the mixing of Markovian channels prompts the question of whether an analogous behavior holds with respect to mixing singular channels. This will be important for understanding the geometry of quantum channels.

 In particular, restricting to the context of mixing Pauli channels, we ask whether singular channels can be produced by mixing non-singular ones, and answer the question in the negative. This negative result implies that non-singular channels form a convex set. On the other hand, mixing singular channels does not necessarily result in a singularity of the resultant channel, showing that singular channels do not form a convex set. Finally, we explain why in the context of mixing Pauli channels, singularities of the channel imply CP-indivisibility, but the converse is not true. This connection between singularity and CP-indivisibility does not hold in general.
\bla

This work is organized as follows. In Sec. \ref{sec:mixcp}, we show that it is not possible to produce singularities of the channel by mixing non-singular channels. In Sec. \ref{sec:type1} and \ref{sec:typeII}, we discuss the results pertaining to mixing singular channels of two broad types. The interplay of singularities and non-Markovianity is discussed in Sec. \ref{sec:xyz}. In all cases, the results are illustrated with examples. Finally, we conclude in Sec. \ref{sec:conc}.

\section{Mixing non-singular Pauli channels \label{sec:mixcp}}

A general Pauli dynamical map is given by
\begin{align}
\mathcal{E}(t) [\rho] = \sum_{i=0}^{3} k_i (t) \sigma_i \rho \sigma^{\dagger}_i ,
\label{eq:paulimap}
\end{align}
where $ \sigma_0 = I $, and $\sigma_i, \; i \in \{1,2,3\}$ are Pauli X, Y, Z operators respectively, and $ \sum_{i=0}^{3} k_i (t) = 1$. The canonical form of master equation corresponding to the map (\ref{eq:paulimap}) has the form 
\begin{align}
\dot{\rho}(t) = \mathcal{L}(t) [\rho(t)] =\sum_{j=1}^{3} \gamma_j(t) (\sigma_j \rho(t) \sigma_j^\dagger - \rho(t))
\label{eq:mastereq}
\end{align}
where $\gamma_j(t)$ are the rates.

 The decay rates may be readily obtained using Eq. (\ref{eq:E2L}) \cite{chruscinski2013non}. Noting that
\begin{align}
	\mathcal{E}(t) [\sigma_j] = \lambda_j(t) \sigma_j.
	\label{eq:pauli-eigen}
\end{align} 
we have
\begin{align}
\dot{\mathcal{E}}(t)[\sigma_j]&= \dot{\lambda}_j(t)\sigma_j \;\; ; \;\;  \mathcal{E}^{-1}(t)[\sigma_j] = \frac{1}{\lambda_j}\sigma_j, \label{eq:eigens}
\end{align} 
showing that the vanishing of a $\lambda_j$ at some time $t_\ast$ corresponds to non-invertibility of the map at that instant, and thus to a singularity of the Pauli channel, per the argument following Eq. (\ref{eq:E2L}). This is made more explicit below.

Now, from (\ref{eq:eigens}) and (\ref{eq:E2L}), we readily obtain the rates in the master equation Eq. (\ref{eq:mastereq}):
\begin{align}
	\gamma_1(t) &= \frac{1}{4} \bigg( \frac{\dot{\lambda}_1(t)}{\lambda_1(t)} - \frac{\dot{\lambda}_2(t)}{\lambda_2(t)} - \frac{\dot{\lambda}_3(t)}{\lambda_3(t)}  \bigg), \nonumber\\ 
	\gamma_2(t) &= \frac{1}{4} \bigg(\frac{\dot{\lambda}_2(t)}{\lambda_2(t)} -\frac{\dot{\lambda}_1(t)}{\lambda_1(t)} - \frac{\dot{\lambda}_3(t)}{\lambda_3(t)}  \bigg), \nonumber\\
	\gamma_3(t) &= \frac{1}{4} \bigg(\frac{\dot{\lambda}_3(t)}{\lambda_3(t)}  - \frac{\dot{\lambda}_1(t)}{\lambda_1(t)} - \frac{\dot{\lambda}_2(t)}{\lambda_2(t)} \bigg). 
	\label{eq:time-dep_rates}
\end{align} 
Note in particular that the $\gamma_j$'s have a singular point when any of the $\lambda_i$ vanishes\bla.  
Our first result, below, essentially asserts the convexity of non-singular Pauli channels.

\textit{Lemma 1.} It is impossible to produce a singular Pauli channel by mixing only non-singular Pauli channels.

\textit{Proof.} Let the Pauli channels that are being mixed be given by
\begin{align}
\mathcal{E}_1(\rho) &\equiv (1-p(t)) \rho(0) + p(t)\sigma_1\rho \sigma_1 , \nonumber \\
\mathcal{E}_2(\rho) &\equiv (1-q(t)) \rho(0) + q(t)\sigma_2\rho \sigma_2, \nonumber \\
\mathcal{E}_3(\rho) &\equiv (1-r(t))\rho(0) +r(t) \sigma_3\rho \sigma_3,
\label{eq:3gamma}
\end{align} 
where the functions $p, q,$ and $r$ quantify the degree of decoherence of the channels and must satisfy $0 \le p, q, r \le 1$ to ensure complete positivity of the maps. The corresponding individual Lindblad rates are
\begin{equation}
\gamma_\eta = \frac{-\dot{\eta}}{1-2\eta},
\label{eq:gammax}
\end{equation}
where $\eta\in \{p(t),q(t),r(t)\}$. Let the three channels in Eq. (\ref{eq:3gamma}) be mixed with probabilities $a, b$ and $c$, where $0 \le a, b, c \le 1$ and $a+b+c=1$. This gives rise to the channel:
\begin{align}
\tilde{\mathcal{E}}(\rho) &= a\mathcal{E}_1(\rho)  + b\mathcal{E}_2(\rho)  + c\mathcal{E}_3(\rho)  \nonumber \\
&=  (1-ap - bq - cr)\rho + ap\sigma_1\rho\sigma_1 + bq \sigma_2\rho\sigma_2 \nonumber \\
&+ cr \sigma_3 \rho \sigma_3. 
\label{eq:this}
\end{align}
By assumption, the mixing maps $\mathcal{E}_1, \mathcal{E}_2$ and $\mathcal{E}_3$ are non-singular. In view of Eq. (\ref{eq:gammax}),  this implies that 
\begin{equation}
0 \le p(t),q(t),r(t) < \frac{1}{2}
\label{eq:pqr}
\end{equation}
for finite time $t$. The time-dependent eigenvalues of the map $\tilde{\mathcal{E}}$ from Eq. (\ref{eq:this}) read
\begin{subequations}
	\begin{align}
	\lambda_1(t) &=1 -2(b q + c r), \label{eq:generala} \\
	\lambda_2(t) &= 1-2(a p + c r),  \\
	\lambda_3(t) &= 1-2(a p+ b q).
	\end{align}
	\label{eq:generalrates}
\end{subequations}
The condition for a singularity in the resultant channel is that one or more of $\lambda_j$ in Eq. (\ref{eq:generalrates}) should vanish at a certain finite time(s) $t_s$. For example, consider $\lambda_1$ in Eq. (\ref{eq:generala}). Given the range restriction Eq. (\ref{eq:pqr}) on the decoherence functions $p(t)$ and $q(t)$, we have 
\begin{equation}
\lambda_1(t) > 1 - (b+c) \ge 0
\label{eq:lambda}
\end{equation}
for finite $t$. Repeating the argument for $\lambda_2$ and $\lambda_3$, we conclude that there can be no singularity in the mixed channel. \hfill $\blacksquare$
\bigskip

It follows from Eq. (\ref{eq:lambda}) and analogous results for $\lambda_2$ and $\lambda_3$ that the non-singular mixing channels necessarily have positive decay rates $\gamma_j$, and therefore are CP-divisible. Thus, as a corollary of Lemma 1, we find that it is impossible to produce a singularity by mixing CP-divisible Pauli channels. 

Lemma 1 does not address the question of whether mixing singular channels produces a singularity in the resultant channel. To address this question, it is convenient to distinguish two types of singular channels. It is clear from Eq. (\ref{eq:gammax}) that for a Pauli channel to be singular, the decoherence function $p(t), q(t)$ or $r(t)$, as the case may be, should attain the value of $\frac{1}{2}$ at some finite time $t$. Accordingly, the two types of Pauli singular channels are those where the value $\frac{1}{2}$ is the maximum or is exceeded. It turns out that they evince quite different behaviors under mixing.

\textit{Definition 2.} Channels of Type I: Those in which the maximum value attained by the decoherence function $p(t), q(t)$ or $r(t)$ in Eq. (\ref{eq:3gamma}) is $\frac{1}{2}$.

In this case, the occurrence of non-Markovianity (CP-indivisibility) can be attributed to the non-monotonicity of the decoherence functions $p(t)$ etc.,  leading to recoherence in the negative slope region of the functions.
Typical instances of interest here would be channels for which the decoherence function is non-monotonic, oscillating between $0$ and $\frac{1}{2}$. In Sec. \ref{sec:type1}, we shall show that: (a) mixing two such channels with singular points $t_\ast^p$ and $t_\ast^q$ produces a singularity only if their singularities are simultaneous ($t_\ast^p = t_\ast^q$), and moreover the resultant singularity occurs at the same time; and furthermore,  (b) three-way mixing of such channels can never produce a singularity.

\textit{Definition 3.} Channels of type II: Those in which the maximum value attained by the decoherence function $p(t), q(t)$ or $r(t)$ in Eq. (\ref{eq:3gamma}) can exceed $\frac{1}{2}$.

Typical instances of interest here would be channels for which the decoherence function is monotonic, reaching an asymptotic value in the interval $(\frac{1}{2},1]$. The positive slope region of the decoherence function $p(t)$ etc., when they exceed half, corresponds to recoherence of the system, leading to non-Markovianity. Unlike in the case of Type I channels, we will find, in Sec. \ref{sec:typeII}, that the features (a) and (b) do not hold in this case, i.e., singularities need not be simultaneous, and the restriction to two-way mixing is not needed.\bla
 
We shall find below that the conditions under which mixing of channels leads to a singularity in the resultant channel depends on the type of the channels being mixed.


We may note that Type I is a more usual occurrence, and can for example result when a qubit system and its qubit environment evolve according to a Hamiltonian given by $\omega(\ket{01}\bra{10} + \ket{10}\bra{01})$ acting on the initial state $\ket{01}$, where $\omega$ is a real number. The joint system remains in the subspace spanned by $\{\ket{01}, \ket{10}\}$, and the reduced state of the system is $\cos^2(\omega t)\ket{0}\bra{0} + \sin^2(\omega t)\ket{1}\bra1$.

\section{ Mixing of Type I channels \label{sec:type1}}


In Eq. (\ref{eq:generalrates}), suppose $\lambda_1(t)$ vanishes at some finite time $t_\ast$. Observe that this can happen only if $q(t_\ast) = r(t_\ast) = \frac{1}{2}$. In other words, the mixing channels $\mathcal{E}_2$ and $\mathcal{E}_3$ must each possess a singularity such that these singularities occur simultaneously at $t_\ast$, which coincides with the singularity in the resultant channel. Moreover, we require that the mixing parameter $a=0$, meaning that only two of three channels should be mixed.  A similar argument holds for $\lambda_2(t)$ and $\lambda_3(t)$.  

To summarize, in the context of mixing Type I channels to produce a singularity,  precisely two channels should be mixed, and they should be synchronized in the occurrence of their singularities. If they are not synchronized, then the singularity will be eliminated in the resultant channel.

The fact that the mixing of two singular channels can produce a non-singular channel can be compared to the situation that mixing CP-indivisible channels can result in a  CP-divisible channel - even a semigroup \cite{wudarski2016markovian, jagadish2020convex}. A consequence is that, like CP-indivisible channels, singular ones also form a nonconvex set.\\ \\
 \textit{Example 1.} Let the mixing channels be $\mathcal{E}_1$ and $\mathcal{E}_2$, with $p(t) = \frac{1}{2}[1-\cos^2(\mu t) ]$ and $q(t) = \frac{1}{2}[1-\cos^2(\nu t)]$, $a, b > 0$ and $a+b=1$ in Eq. (\ref{eq:this}). In view of Eq. (\ref{eq:time-dep_rates}), the eigenvalues $\lambda_i(t)$ of the resultant channel read
\begin{align}
\lambda_1(t) &=1 - b \sin^2 (\nu t),  \nonumber \\
\lambda_2(t) &= 1- a \sin^2 (\mu t), \nonumber \\
\lambda_3(t) &= 1- a \sin^2 (\mu t) - b \sin^2 (\nu t),
\end{align}
showing that there is a singularity  only from the zeros of $\lambda_3(t)$, and furthermore this happens if and only if the two trigonometric terms attain 1 at the same time $t_\ast$, which will also be the singular point of the resultant channel. A simple way to ensure this is by having $\mu = \nu$, in which case singularities occur in the resultant channel for $t = n\frac{\pi}{2}$.\hfill $\blacklozenge$\\
\bigskip

It is not hard to show that this behavior, of the singularities of the mixing channels to be simultaneous at some time $t_\ast$, and leading to a singularity at the same time $t_\ast^R = t_\ast$ in the resultant channel, is general for Type I channels.

The following example illustrates the idea that the number of mixing channels should not exceed 2. Otherwise the singularity is eliminated.\\ \\
\textit{Example 2.} A depolarizing colored noise is the Random telegraph noise (RTN)  non-Markovian depolarizing channel $\mathcal{E}[\rho] = \sum_i A_i \rho A_i^\dagger$, with the Kraus operators \cite{daffer2004depolarizing}
$
A_i = \sqrt{P_i} \sigma_i,
$
where $\sigma_0=I, \sigma_x=\sigma_1,\sigma_y=\sigma_2,\sigma_z=\sigma_3$ are Pauli operators. Here,
\begin{align}
P_0 &= \frac{1}{4}[1 + \Lambda_1 + \Lambda_2 + \Lambda_3],\nonumber \\
P_1 &= \frac{1}{4}[1 + \Lambda_1 - \Lambda_2 - \Lambda_3], \nonumber \\
P_2 &= \frac{1}{4}[1 - \Lambda_1 + \Lambda_2 - \Lambda_3], \nonumber \\
P_3 &= \frac{1}{4}[1 - \Lambda_1 - \Lambda_2 + \Lambda_3], 
\label{eq:RTN}
\end{align} 
where 
\begin{align}
\Lambda_i = \exp (-w t) \left[\frac{\sin (w t \mu_i)}{\mu_i}+\cos (w t \mu_i)\right],
\label{eq:Lambda}
\end{align}
The quantity $w = \frac{1}{2 \tau}$ is the spectral bandwidth while $\tau$ is the rate of fluctuation of the environment affecting the qubit, and $\mu_i = \sqrt{\left(\frac{2 d_i}{w}\right)^2-1}$, with $d_{i}$ representing the coupling strengths corresponding to the $i$th Pauli channel. For the present purpose, let all $d_i$'s in Eq. (\ref{eq:Lambda}) be taken to be equal, given by $d$. This corresponds to equal mixing of the $X,Y,$ and $Z$ Pauli RTN channels, as a result of which we obtain an isotropic RTN Pauli channel. 



Now consider individual RTN Pauli channels of Type I with their respective decoherence function being
\begin{align}
p(t) = q(t)=r(t) = \frac{1- \Lambda(t)}{2}, 
\label{eq:RTNpqr}
\end{align}
where $\Lambda(t)$ is given by Eq. (\ref{eq:Lambda})  with $\mu_1=\mu_2=\mu_3$. 
We now consider the question of whether the above RTN  non-Markovian depolarizing channel can be reproduced by mixing the individual RTN Pauli channels of Type I. For $d \gg w$, the zeros of $\Lambda$ occur periodically, making the channel possess an infinite number of singularities. For $d < w$, $\Lambda(t)$ attains zero only at $t=\infty$, making the channel non-singular. 

Eq. (\ref{eq:generalrates}) yields the following eigenvalues of the resultant channel:
\begin{align}
\lambda_1(t) &=1 - (b+c) (1-\Lambda),  \nonumber\\
\lambda_2(t) &= 1 -(a+c) (1-\Lambda), \nonumber\\
\lambda_3(t) &= 1-(a + b )(1-\Lambda).
\label{eq:RTNlambdas2}
\end{align}
Whilst in general $\Lambda$ takes values in the range $(-1,+1]$, but to conform to Type I, the parameters $d$ and $w$ must be so chosen that $\Lambda(t)$ is confined in the range $[0,1]$, with the singularity occurring when $\Lambda(t)=0$. If $a, b, c > 0$, then the sum of any two of them is strictly less than 1. It follows from Eq. (\ref{eq:RTNlambdas2}) and the Type I restriction (requiring that $\Lambda$ is bounded below by 0) that each $\lambda_j$ in Eq. (\ref{eq:RTNlambdas2}) never vanishes.  \hfill $\blacklozenge$

We now consider an analogous result when type II channels are mixed, and show how it contrasts with the above two examples.

\section{Mixing of Type II channels \label{sec:typeII}}

In Equation (\ref{eq:generalrates}), if we relax the requirement that $p, q$ and $r$ are bounded by $\frac{1}{2}$, then we obtain Type II channels.  In this case, we will find that neither the synchronization nor restriction of the mixing channels to two, is required, for producing a singular channel. \\ \\
\textit{Example 3.} Consider the same system as in Example 2, but letting $p, q,$ and $r$ to exceed $\frac{1}{2}$. Accordingly, $\Lambda$ takes values in the range $(-1,+1]$. For simplicity, let $a=b=c=\frac{1}{3}$, in which case the eigenvalues become 
\begin{align}
\lambda_{1,2,3} = \frac{1}{3}[1+ 2 \Lambda(t)].
\end{align}
The singularity of the resulting RTN depolarizing channel occurs when $\Lambda(t)=-\frac{1}{2}.$ From Eq. (\ref{eq:RTNpqr}), we find that this singularity happens when $p(t)=q(t)=r(t) = \frac{3}{4}$, which is permissible for mixing channels of type II. \hfill $\blacklozenge$ 

As a final example, consider the three mixing channels to be of type II, having possibly different functional forms, but all reaching an asymptotic value greater than (say) $\frac{4}{5}$. \\ \\
\textit{Example 4.}  At some time $t_\ast^R$, let $q(t_\ast^R) = \frac{3}{5}$ and $r(t_\ast^R) = \frac{4}{5}$. The singularities of $\mathcal{E}_Y$ and $\mathcal{E}_Z$ occur, respectively, at $t_\ast^2$ and $t_\ast^3$, where $q(t_\ast^2) = r(t_\ast^3) = \frac{1}{2}$, where in general we don't require $t^2_\ast = t^3_\ast$, i.e., the singularities of the mixing channels are not necessarily synchronized. Furthermore let the mixing fraction $b=\frac{1}{6}$ and $c=\frac{1}{2}$, so that $a = 1-b-c=\frac{1}{3}$, implying that there is a finite fraction of the channel $\mathcal{E}_X$ in the mixing. It follows from Eq. (\ref{eq:generala}) that $\lambda_1=0$ at $t_\ast^R$, meaning that this is a singularity of the resultant channel. Note that $t^R_\ast$ need not coincide with either $t^j_\ast$ $(j = 2,3)$. \hfill$\blacklozenge$



\section{Interplay of singularities and non-Markovianity\label{sec:xyz}}

It turns out that for the resultant channels considered here, singular channels are necessarily non-Markovian (in the CP-indivisible sense). In Eq. (\ref{eq:generalrates}), consider the point $t_\ast$ where the first singularity is encountered, i.e., one of the $\lambda_j(t_\ast)$ vanish, say $\lambda_1(t_\ast)=0$. From Eq. (\ref{eq:generala}), we have $\dot{\lambda}_1(t) \equiv -2(b\dot{q} + c\dot{r})$. In the case of Type I channels, both $q(t)$ and $r(t)$ reach $\frac{1}{2}$ and fall off at the same time. Therefore, $\dot{\lambda}_1(t)$  is negative just before $t_\ast$, and flips sign to positive just after $t_\ast$. On the other hand, $\lambda_1$ remains positive for all time. Thus:
\begin{eqnarray}
\lim_{t \rightarrow t_\ast^\pm} \frac{\dot{\lambda_1}}{\lambda_1} = \pm \infty.
\label{eq:infty+}
\end{eqnarray}
This implies that $\gamma_2$ and $\gamma_3$ flip the sign from positive to negative at the singularity, and $\gamma_1$ the other way.


For Type II channels, by virtue of monotonic increase of $q(t)$ and $r(t)$, $\dot{\lambda}_1(t) \equiv -2(b\dot{q} + c\dot{r}) < 0$ for all time. On the other hand, $\lambda_1$ flips its sign from positive to negative at the singularity. Thus, Eqs. (\ref{eq:infty+}) and the attendant consequences for the decay rates hold here too. Therefore, curiously, despite the contrasting behavior in the eigenvalues and the rate of change, yet in both Type I and Type II channels, singularities signal CP-indivisibility in a similar way.

It may be worth pointing out that singularities do not necessarily imply CP-indivisibility. For illustration, consider a CP-indivisible dephasing channel described by $\mathcal{L}[\rho] \equiv \gamma(t)(\sigma_3 \rho \sigma_3-\rho)$ with the decay rate $\gamma(t) = \tan(\omega t)$, where $ \omega $ is some real number. The channel has an infinitely many singularities at $t_\ast = \frac{(2n+1)\pi}{2 \omega}$ for integer $n$, which will flip the sign of the rate, and thus signal CP-indivisibility in a similar manner as discussed above. 
By contrast, consider the same channel, but with decay rate $\gamma(t) = \tan^2(\omega t)$. This channel contains singularities at the same instants as the above channel, but the sign of the rate never flips from positive to negative at any of these singularities and thus corresponds to a CP-divisible process. 

\section{Discussions and Conclusions \label{sec:conc}}

This work discusses the problem of producing a singular general Pauli dynamical map by the mixing  non-singular (or,  regular) Pauli channels. We point out that it is impossible to do so. Different conditions on the classes of mixing singular channels are considered in order to guarantee that the resultant channel is singular.  In particular, we show that: (i) for Type I channels it is possible to produce a singular channel by mixing two singular Pauli channels, provided the occurrence of their singularities is synchronized; (ii) mixing three Type I singular channels results in the elimination of singularities i.e., such a convex combination results in a regular channel; and (iii) by contrast, in the case of Type II channels, we have shown that mixing two or three singular channels can result in a singular channel, and  the singularities need not be synchronized in their occurrence. 

 A further question that may be considered here is the power of mixing weaker forms of non-Markovianity than CP-indivisibility (cf. Reference \cite{shrikant2020temporal} and references therein), in terms of generating singularities and/or stronger forms of non-Markovianity. An aspect of this for the measure of CP-indivisible maps produced by mixing a class of Pauli maps was considered in Reference \cite{jagadish2020+convex}. A future direction would be to explore the results of this paper from a geometric point of view \cite{jagadish2019+measure,jagadish2019measure,siudzinska2019geometry}, in particular to quantify the measure of non-singular channels produced by mixing singular ones, analogous to results for CP-divisible channels in References \cite{jagadish2020convex,jagadish2020+convex}. Reference \cite{megier2020interplay} shows how an operation of coarse-graining in time while transforming a master equation from a nonlocal integro-differential form to a time-local one can modify the CP-indivisibility property of the dynamics \cite{megier2020interplay}. An interesting question here would be whether this approximation procedure can also modify the (non-)singular property of the dynamics. 

 Ref. \cite{uriri2020experimental} reports on an experimental implementation of producing non-Markovianity by two-way mixing of Pauli semigroups in a linear quantum optical platform, an extension of which to three-way mixing of more general CP-divisible maps has recently been proposed \cite{jagadish2020+convex}. This idea may be adapted for mixing singular channels, which can be produced by suitable bath engineering, possibly with appropriate modifications to the experimental implementation in Ref. \cite{cialdi2019experimental} in the case of a photonic realization, or in Reference \cite{carmele2019non}, in the case of a semiconductor quantum optical realization. \bla 
 
 Our result here, that singularities cannot be produced by mixing non-singular channels, is shown only for qubit Pauli channels. We expect that this is true quite generally, and in particular, applicable to non-Pauli maps. 

\acknowledgements

SU and VNR acknowledge Admar Mutt Education Foundation for support from a scholarship. RS acknowledges the support of Department of Science and Technology (DST), India, under Grant No. MTR/2019/001516. RS and SB acknowledge, respectively, the support from the Interdisciplinary Cyber Physical Systems (ICPS) program of the Department of Science and Technology (DST), India, under Grants No.: DST/ICPS/QuEST/Theme-1/2019/14 and No. DST/ICPS/QuEST/Theme-1/2019/6.

\bibliography{mixing}

\begin{thebibliography}{10}

\bibitem{banerjee2018open}
Subhashish Banerjee.
\newblock {\em Open quantum systems : Dynamics of Nonclassical Evolution},
  volume~20.
\newblock Springer Nature Singapore, 2018.

\bibitem{breuer2002theory}
Heinz-Peter Breuer and Francesco Petruccione.
\newblock {\em The theory of open quantum systems}.
\newblock Oxford University Press, 2002.

\bibitem{srikanth2008squeezed}
R~Srikanth and Subhashish Banerjee.
\newblock Squeezed generalized amplitude damping channel.
\newblock {\em Phys. Rev. A}, 77(1):012318, 2008.

\bibitem{omkar2013dissipative}
S~Omkar, R~Srikanth, and Subhashish Banerjee.
\newblock Dissipative and non-dissipative single-qubit channels: dynamics and
  geometry.
\newblock {\em Qu. Inf. Proc}, 12(12):3725--3744, 2013.

\bibitem{RHP10}
{\'A}ngel Rivas, Susana~F Huelga, and Martin~B Plenio.
\newblock Entanglement and non-markovianity of quantum evolutions.
\newblock {\em Phys. Rev. Lett}, 105(5):050403, 2010.

\bibitem{breuer2009measure}
Heinz-Peter Breuer, Elsi-Mari Laine, and Jyrki Piilo.
\newblock Measure for the degree of non-markovian behavior of quantum processes
  in open systems.
\newblock {\em Phys. Rev. Lett}, 103(21):210401, 2009.

\bibitem{hall2014canonical}
Michael J.~W. Hall, James~D. Cresser, Li~Li, and Erika Andersson.
\newblock Canonical form of master equations and characterization of
  non-markovianity.
\newblock {\em Phys. Rev. A}, 89:042120, Apr 2014.

\bibitem{shrikant2020temporal}
Shrikant Utagi, R~Srikanth, and Subhashish Banerjee.
\newblock Temporal self-similarity of quantum dynamical maps as a concept of
  memorylessness.
\newblock {\em Scientific Reports}, 10(1):1--10, 2020.

\bibitem{li2019non}
C-F Li, G-C Guo, and J~Piilo.
\newblock Non-{M}arkovian quantum dynamics: What does it mean?
\newblock {\em EPL (Europhysics Letters)}, 127(5):50001, 2019.

\bibitem{verstraete2002quantum}
Frank Verstraete and Henri Verschelde.
\newblock On quantum channels.
\newblock {\em arXiv:0202124}, 2002.

\bibitem{ruskai2002analysis}
Mary~Beth Ruskai, Stanislaw Szarek, and Elisabeth Werner.
\newblock An analysis of completely-positive trace-preserving maps on m2.
\newblock {\em Linear algebra and its applications}, 347(1-3):159--187, 2002.

\bibitem{banerjee2007dynamics}
Subhashish Banerjee and R~Ghosh.
\newblock Dynamics of decoherence without dissipation in a squeezed thermal
  bath.
\newblock {\em Journal of Physics A: Mathematical and Theoretical},
  40(45):13735, 2007.

\bibitem{naikoo2020coherence}
Javid Naikoo and Subhashish Banerjee.
\newblock Coherence-based measure of quantumness in (non-) markovian channels.
\newblock {\em Quantum Information Processing}, 19(1):29, 2020.

\bibitem{esposito2009nonequilibrium}
Massimiliano Esposito, Upendra Harbola, and Shaul Mukamel.
\newblock Nonequilibrium fluctuations, fluctuation theorems, and counting
  statistics in quantum systems.
\newblock {\em Reviews of modern physics}, 81(4):1665, 2009.

\bibitem{campisi2011colloquium}
Michele Campisi, Peter H{\"a}nggi, and Peter Talkner.
\newblock Colloquium: Quantum fluctuation relations: Foundations and
  applications.
\newblock {\em Reviews of Modern Physics}, 83(3):771, 2011.

\bibitem{thomas2018thermodynamics}
George Thomas, Nana Siddharth, Subhashish Banerjee, and Sibasish Ghosh.
\newblock Thermodynamics of non-markovian reservoirs and heat engines.
\newblock {\em Phys. Rev. E}, 97:062108, Jun 2018.

\bibitem{thapliyal2017quantum}
Kishore Thapliyal, Anirban Pathak, and Subhashish Banerjee.
\newblock Quantum cryptography over non-markovian channels.
\newblock {\em Quantum Information Processing}, 16(5):115, 2017.

\bibitem{shrikant2020ping}
Shrikant Utagi, R~Srikanth, and Subashish Banerjee.
\newblock Ping-pong quantum key distribution with trusted noise: Non-Markovian
  advantage.
\newblock Quantum Inf.Process.19,366 (2020).

\bibitem{kumar2018non}
N~Pradeep Kumar, Subhashish Banerjee, R~Srikanth, Vinayak Jagadish, and
  Francesco Petruccione.
\newblock Non-markovian evolution: a quantum walk perspective.
\newblock {\em Open Systems \& Information Dynamics}, 25(03):1850014, 2018.

\bibitem{naikoo2020non}
Javid Naikoo, Subhashish Banerjee, and CM~Chandrashekar.
\newblock Non-markovian channel from the reduced dynamics of coin in quantum
  walk.
\newblock Physical Review A, 102(6), 062209 (2020).

\bibitem{bhattacharya2018evolution}
Samyadeb Bhattacharya, Subhashish Banerjee, and Arun~Kumar Pati.
\newblock Evolution of coherence and non-classicality under global
  environmental interaction.
\newblock {\em Quantum Information Processing}, 17(9):236, 2018.

\bibitem{banerjee2010dynamics}
Subhashish Banerjee, V~Ravishankar, and R~Srikanth.
\newblock Dynamics of entanglement in two-qubit open system interacting with a
  squeezed thermal bath via dissipative interaction.
\newblock {\em Annals of Physics}, 325(4):816--834, 2010.

\bibitem{naikoo2019facets}
Javid Naikoo, Supriyo Dutta, and Subhashish Banerjee.
\newblock Facets of quantum information under non-markovian evolution.
\newblock {\em Phys. Rev. A}, 99:042128, Apr 2019.

\bibitem{li2020non}
C-F Li, G-C Guo, and J~Piilo.
\newblock Non-{M}arkovian quantum dynamics: What is it good for?
\newblock {\em EPL (Europhysics Letters)}, 128(3):30001, 2020.

\bibitem{gorini1976completely}
Vittorio Gorini, Andrzej Kossakowski, and Ennackal Chandy~George Sudarshan.
\newblock Completely positive dynamical semigroups of n-level systems.
\newblock {\em Journal of Mathematical Physics}, 17(5):821--825, 1976.

\bibitem{lindblad1975completely}
G{\"o}ran Lindblad.
\newblock Completely positive maps and entropy inequalities.
\newblock {\em Communications in Mathematical Physics}, 40(2):147--151, 1975.

\bibitem{chruscinski2010non}
Dariusz Chru{\'s}ci{\'n}ski and Andrzej Kossakowski.
\newblock Non-markovian quantum dynamics: local versus nonlocal.
\newblock {\em Physical review letters}, 104(7):070406, 2010.

\bibitem{chruscinski2011divisiblity}
Dariusz Chru{\'s}ci{\'n}ski, Andrzej Kossakowski, and {\'A}ngel Rivas.
\newblock Measures of non-markovianity: Divisibility versus backflow of
  information.
\newblock {\em Physical Review A}, 83(5):052128, 2011.

\bibitem{RHP14}
Angel Rivas, Susana~F Huelga, and Martin~B Plenio.
\newblock Quantum non-markovianity: characterization, quantification and
  detection.
\newblock {\em Rep. Prog. Phys}, 77(9):094001, 2014.

\bibitem{chruscinski2014degree}
Dariusz Chru{\'s}ci{\'n}ski and Sabrina Maniscalco.
\newblock Degree of non-markovianity of quantum evolution.
\newblock {\em Physical review letters}, 112(12):120404, 2014.

\bibitem{bylicka2014non}
B~Bylicka, D~Chru{\'s}ci{\'n}ski, and Sci Maniscalco.
\newblock Non-markovianity and reservoir memory of quantum channels: a quantum
  information theory perspective.
\newblock {\em Sci. Rep}, 4,1 (2014).

\bibitem{li2018concepts}
Li~Li, Michael~J.W. Hall, and Howard~M. Wiseman.
\newblock Concepts of quantum non-markovianity: A hierarchy.
\newblock {\em Physics Reports}, 759:1 -- 51, 2018.

\bibitem{shrikant2021causality}
Shrikant Utagi.
\newblock Quantum causal correlations and non-markovianity of quantum
  evolution.
\newblock {\em Physics Letters A}, 386:126983, 2021.

\bibitem{hou2012singularity}
SC~Hou, XX~Yi, SX~Yu, and CH~Oh.
\newblock Singularity of dynamical maps.
\newblock {\em Physical Review A}, 86(1):012101, 2012.

\bibitem{shrikant2018non-Markovian}
U.~Shrikant, R.~Srikanth, and Subhashish Banerjee.
\newblock Non-markovian dephasing and depolarizing channels.
\newblock {\em Phys. Rev. A}, 98:032328, Sep 2018.

\bibitem{hou2011alternative}
SC~Hou, XX~Yi, SX~Yu, and CH~Oh.
\newblock Alternative non-markovianity measure by divisibility of dynamical
  maps.
\newblock {\em Physical Review A}, 83(6):062115, 2011.

\bibitem{wolf2008assessing}
Michael~Marc Wolf, J~Eisert, TS~Cubitt, and J~Ignacio Cirac.
\newblock Assessing non-markovian quantum dynamics.
\newblock {\em Physical review letters}, 101(15):150402, 2008.

\bibitem{chruscinski2010long}
Dariusz Chru{\'s}ci{\'n}ski, Andrzej Kossakowski, and Saverio Pascazio.
\newblock Long-time memory in non-markovian evolutions.
\newblock {\em Physical Review A}, 81(3):032101, 2010.

\bibitem{wudarski2016markovian}
Filip~A Wudarski and Dariusz Chru{\'s}ci{\'n}ski.
\newblock Markovian semigroup from non-markovian evolutions.
\newblock {\em Physical Review A}, 93(4):042120, 2016.

\bibitem{jagadish2020convex}
Vinayak Jagadish, R~Srikanth, and Francesco Petruccione.
\newblock Convex combinations of pauli semigroups: Geometry, measure, and an
  application.
\newblock {\em Physical Review A}, 101(6):062304, 2020.

\bibitem{jagadish2020+convex}
Vinayak Jagadish, R~Srikanth, and Francesco Petruccione.
\newblock Convex combinations of cp-divisible pauli channels that are not
  semigroups.
\newblock {\em Physics Letters A}, 384:126907, 2020.

\bibitem{shrikant2020quasi}
Shrikant Utagi, Vinod~N Rao, R~Srikanth, and Subhashish Banerjee.
\newblock A class of quasi-eternal non-Markovian Pauli channels and their measure.
\newblock {\em arXiv preprint arXiv:2002.11452}, 2020.

\bibitem{chruscinski2013non}
Dariusz Chru{\'s}ci{\'n}ski and Filip~A Wudarski.
\newblock Non-markovian random unitary qubit dynamics.
\newblock {\em Physics Letters A}, 377(21-22):1425--1429, 2013.

\bibitem{daffer2004depolarizing}
Sonja Daffer, Krzysztof W{\'o}dkiewicz, James~D Cresser, and John~K McIver.
\newblock Depolarizing channel as a completely positive map with memory.
\newblock {\em Phys. Rev. A}, 70(1):010304, 2004.

\bibitem{jagadish2019+measure}
Vinayak Jagadish, R~Srikanth, and Francesco Petruccione.
\newblock Measure of positive and not completely positive single-qubit pauli
  maps.
\newblock {\em Physical Review A}, 99(2):022321, 2019.

\bibitem{jagadish2019measure}
Vinayak Jagadish, R.~Srikanth, and Francesco Petruccione.
\newblock Measure of not-completely-positive qubit maps: The general case.
\newblock {\em Phys. Rev. A}, 100:012336, Jul 2019.

\bibitem{siudzinska2019geometry}
Katarzyna Siudzi{\'n}ska.
\newblock Geometry of pauli maps and pauli channels.
\newblock {\em Physical Review A}, 100(6):062331, 2019.

\bibitem{megier2020interplay}
Nina Megier, Andrea Smirne, and Bassano Vacchini.
\newblock The interplay between local and non-local master equations: exact and
  approximated dynamics.
\newblock {\em New Journal of Physics}, 22(8):083011, aug 2020.

\bibitem{uriri2020experimental}
SA~Uriri, F~Wudarski, I~Sinayskiy, F~Petruccione, and MS~Tame.
\newblock Experimental investigation of markovian and non-markovian channel
  addition.
\newblock {\em Physical Review A}, 101(5):052107, 2020.

\bibitem{cialdi2019experimental}
Simone Cialdi, Claudia Benedetti, Dario Tamascelli, Stefano Olivares, Matteo~GA
  Paris, and Bassano Vacchini.
\newblock Experimental investigation of the effect of classical noise on
  quantum non-markovian dynamics.
\newblock {\em Physical Review A}, 100(5):052104, 2019.

\bibitem{carmele2019non}
Alexander Carmele and Stephan Reitzenstein.
\newblock Non-markovian features in semiconductor quantum optics: quantifying
  the role of phonons in experiment and theory.
\newblock {\em Nanophotonics}, 8(5):655--683, 2019.

\end{thebibliography}

\end{document}